\documentclass[12pt]{article}
\usepackage[T1]{fontenc}
\usepackage{geometry}
\usepackage{graphics}
\usepackage{setspace}
\makeatletter

\newcommand{\LyX}{L\kern-.1667em\lower.25em\hbox{Y}\kern-.125emX\@}
\newcommand{\lyxline}[1]{
  {#1 \vspace{1ex} \hrule width \columnwidth \vspace{1ex}}
}

\usepackage{verbatim}

\usepackage{latexsym}
\usepackage{amssymb}
\makeatother

\begin{document}

{\par\centering \textbf{Version Date :} 21\( ^{th} \) of July 1999\par}
\bigskip{}

{\par\centering \textbf{\LARGE The broad Brillouin doublets and central peak
of KTaO\( _{3} \) }\LARGE \par}

\bigskip{}
{\par\centering \textbf{\large E. Farhi\( ^{a,b} \), A.K. Tagantsev\( ^{c} \),
B. Hehlen\( ^{a} \), R. Currat\( ^{b} \), L.A. Boatner\( ^{d} \), and E.
Courtens\( ^{a} \)}\large \par}
\bigskip{}

{\par\centering \textit{\small \( ^{a} \) Laboratoire des Verres, Universit\'e
de Montpellier 2, F-34095 Montpellier, France}\small \par}

{\par\centering \textit{\small \( ^{b} \) Institut Laue Langevin, BP 156, 38042
Grenoble Cedex 9, France }\small \par}

{\par\centering \textit{\small \( ^{c} \) Laboratoire de C\'eramique, EPFL,
CH-1015 Lausanne, Switzerland}\small \par}

{\par\centering \textit{\small \( ^{d} \) Solid State Division, Oak Ridge Nat.
Lab., Oak Ridge, TN 37831-6056, USA}\small \par}

\begin{abstract}
\lyxline{\small}\vspace{-1\parskip}
The incipient ferroelectric KTaO\( _{3} \) presents low-\( T \) Brillouin
spectra anomalies, \emph{e.g.} a broad central peak (CP), and some additional
Brillouin doublets (BD), whose origin is interpreted in terms of phonon-density
fluctuation processes. A parameterisation from new extensive high-resolution
neutron-scattering measurements is used to show that hydrodynamic second sound
from high damping (compared to BD frequency) TA phonons may exist in the crystal.
Furthermore, low damping thermal phonons may scatter light through two-phonon
difference processes and appear on the Brillouin spectra either as a sharp or
a broader BD, depending on the phonon damping and group velocity . The comparison
between computed anisotropies and experimental measurements favours the second
process.\lyxline{\small}

\end{abstract}
\bigskip{}
\textit{keywords} : Computer modelling, Perovskite compounds, Phonons, Spectroscopy.
\bigskip{}

\textbf{Corresponding Author :} E. Farhi

\textbf{Address :} Institut Laue-Langevin, BP 156, F-38042 Grenoble Cedex 9,France

\textbf{Fax :} 33 (0)4 76 48 39 06

\textbf{E-mail Address :} farhi@ill.fr 

\textbf{PACS numbers :} 77.84 Dy, 63.20.-e, 62.20.Dc, 78.70.Nx, 66.90+r, 78.35+c
\bigskip{}

{\par\raggedleft \textit{Physica B Preprint}\par}

\newpage


\onecolumn

Incipient ferroelectrics SrTiO\( _{3} \) or KTaO\( _{3} \) are perovskite
type crystals that remain paraelectric down to the lowest temperature, due to
zero-point quantum fluctuations \cite{Muller79}. The dielectric constant saturates
at a high value as the soft transverse optic (TO) zone-centre frequency stabilises
to a low finite value. Since the case of SrTiO\( _{3} \) is complicated by
the structural transition at \( T_{a}\sim  \) 105 \cite{Courtens96}, we study
KTaO\( _{3} \) which remains in the cubic phase down to the lowest temperatures. 

Experimental evidence of anomalous behaviour in pure KTaO\( _{3} \) when cooling
into the quantum paraelectric (QPE) regime are numerous. In Particular, the
low energy Brillouin spectra present unusual features. Hehlen \textit{et al.}
\cite{Hehlen95b} found a new sound-like Brillouin doublet (BD) both in SrTiO\( _{3} \)
and KTaO\( _{3} \) in addition to the quasi-elastic broad central peak (CP)
first reported by Lyons and Fleury \cite{Lyons76}. It was interpreted as the
signature of second sound \cite{Hehlen95b,Gurevich88}. Our purpose here is
to investigate the origin of CP and BD in the frame of phonon-density fluctuation
processes. 

The anisotropy of anomalous Brillouin contributions was measured in KTaO\( _{3} \)
Ultra High Purity crystals along the three high symmetry axes by means of Brillouin
light-backscattering, with a momentum transfer \( |\vec{q}|= \) 3.55 10\( ^{-3} \)
rlu. The observations of Hehlen \textit{et al.} \cite{Courtens96,Hehlen95b}
are confirmed for the \( \vec{q} \) // <001> symmetry axis. A sharp BD appears
upon the broad CP \cite{Lyons76} below \( T\lesssim  \) 22 K. Its associated
velocity is \( v_{BD}\sim  \) 1100 m/s. In the <110> direction, the BD and
CP contributions are also observed, as well as an additional broad BD (BBD)
for temperatures between 30 and 100 K. While the CP and BBD contributions are
also seen along the <111> direction, no sharp BD could be measured for temperatures
as low as 5 K \cite{Farhi98,PRBsubmit}. 

In order to look for an explanation of CP, BD, and BBD contributions in terms
of phonon-density fluctuation processes, we parameterised the low-energy phonon
dispersion surfaces \( \omega (\vec{q}) \) below 100 K within a simple phenomenological
quasi-harmonic anisotropic model due to Vaks \cite{PRBsubmit,Vaks68}. The precise
determination of the necessary parameters required additional new inelastic
neutron-scattering experiments to be carried out. Taking into account the 4-dimensional
nature of the instrumental resolution ellipsoid and the anisotropy of the dispersion
surfaces, an effective energy uncertainty as low as 0.1 meV was achieved. The
resulting dispersion curves and their parameterisation are presented in Fig.
\ref{DC} along the principal symmetry axes, showing the well-known TO-TA coupling
\cite{Perry89}. The transverse phonon dispersion surfaces show deep \emph{'valleys'}
along the <001> directions and for \( \vec{q} \) // <110 > with \( \vec{e} \)
// <001>. Some \emph{'uphill'} regions are found elsewhere, specially for \( \vec{q} \)
// <111>, and \( \vec{q} \) // <110 > with \( \vec{e} \) // <1\( \overline{1} \)0>.
This anisotropic character was confirmed by off-principal-axis measurements.

We now analyse the origin of the anomalous Brillouin contributions in terms
of classical phonon-kinetics processes. Second sound (2S) is the collective
mode of phonon-density fluctuations induced by multiphonon processes \cite{Gurevich88,Wehner72}.
It may propagate at a frequency \( \omega _{2S} \) in a pure crystal when the
'window' condition, \( \Gamma _{R}<\omega _{2S}<\Gamma _{N} \) (1), is fulfilled.
Here, \( \Gamma _{N} \) and \( \Gamma _{R} \) are the normal and resistive
damping coefficients respectively. In KTaO\( _{3} \), Eq. (1) is satisfied
since the proximity of the optic and acoustic modes enhances the relaxation
processes. The computation of 2S velocity using Vaks model and the thermalized
part of the phonon spectrum gives \( v_{2S} \) = 1100 \( \pm  \) 200 m/s,
in excellent agreement with measurements of BD velocity. However, it seems difficult
to explain the directional dependence of BD in this framework.

When Eq. (1) is not satisfied, low-damping phonons might interact \emph{via}
two-phonon difference scattering (TPDS). Following Wehner and Klein \cite{Farhi98,Wehner72},
we evaluated the corresponding Brillouin spectra for different anisotropic dispersion
surfaces. It was found that TPDS from low-\( \Gamma _{N} \) \emph{valley}-TA
with small group velocity \( v_{g} \) appears as a sharp BD on low-\( T \)
spectra, whereas a BBD is obtained by TPDS at slightly higher \( T \) from
thermalized TA phonons around \emph{uphill} regions, where \( \Gamma _{N} \)
and \( v_{g} \) are higher. Finally, a broad CP originates from high-\( \Gamma _{N} \)
TO phonon TPDS. Thus, the intrinsically anisotropic TPDS spectral shapes are
in agreement with Brillouin experiments.

As a conclusion, we reported new Brillouin spectroscopy and high-resolution
inelastic neutron-scattering results in pure KTaO\( _{3} \). On the basis of
those experimental data, a simple phenomenological parameterisation of the phonon
dispersion surfaces was used to study two different scenarios for the appearance
of additional Brillouin contributions : two-phonon difference scattering processes
(TPDS) and second sound (2S). The comparison between computed an\-isotropies
and experimental measurements seems to favour the TPDS explanation.

\newpage

\onecolumn
\begin{figure}

\caption{\label{DC}Low-\protect\( T\protect \) dispersion curves and parameterisation
of KTaO\protect\( _{3}\protect \) low energy phonons. The mean energy uncertainty
is 0.1 meV. The LA mode is for \protect\( \vec{q}\protect \) = <0 0 2+\protect\( \xi \protect \)>.}
\end{figure}
\newpage

\onecolumn

{\par\centering \includegraphics{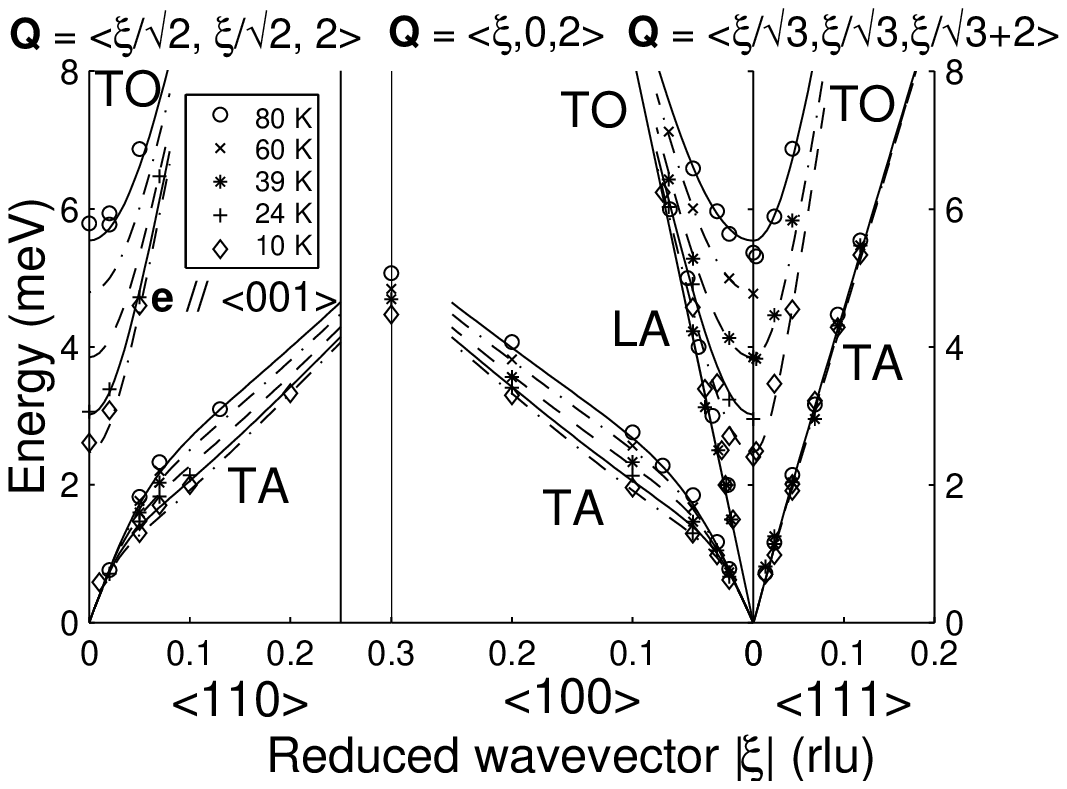} \par}

\end{document}